\documentclass[a4paper,twoside]{amsart}
\usepackage{amsmath}
\usepackage{amsfonts}
\usepackage{amssymb}
\usepackage{amsthm}
\usepackage{newlfont}
\usepackage{graphicx}
\usepackage{amscd}

\theoremstyle{plain}
\newtheorem{Th}{Theorem}[section]
\newtheorem{Cor}[Th]{Corollary}
\newtheorem{Lem}[Th]{Lemma}
\newtheorem{Prop}[Th]{Proposition}
\theoremstyle{definition}
\newtheorem{Def}{Definition}[section]

\theoremstyle{remark}
\newtheorem*{Rem}{Remark}
\numberwithin{equation}{section}
\newcommand{\PP}{{\mathbb P}}

\newcommand{\GG}{{\mathbb G}}
\newcommand{\CC}{{\mathbb C}}
\newcommand{\DD}{{\mathbb D}}

\newcommand{\ZZ}{{\mathbb Z}}

\newcommand{\RR}{{\mathbb R}}

\newcommand{\cL}{{\mathcal L}}

\newcommand{\bphi}{\boldsymbol{\phi}}
\newcommand{\bpsi}{\boldsymbol{\psi}}

\begin{document}

\title
{Desargues maps and the Hirota--Miwa equation}

\author{Adam Doliwa}

\address{Faculty of Mathematics and Computer Science, University of Warmia and
Mazury,
ul.~\.{Z}o{\l}\-nierska~14, 10-561 Olsztyn, Poland}

\email{doliwa@matman.uwm.edu.pl}

\date{}

\begin{abstract}
We study the Desargues maps $\phi:\ZZ^N\to\PP^M$, which generate lattices
whose points are collinear with all their nearest (in positive 
directions) neighbours. The  
multidimensional compatibility of the map is equivalent to the Desargues
theorem and its higher-dimensional generalizations. 
The nonlinear counterpart of the map is the non-commutative (in general)
Hirota--Miwa system. In the commutative
case of the complex field we apply the
nonlocal $\bar\partial$-dressing method to construct Desargues maps and
the corresponding solutions of the system. 
In particular, we identify the Fredholm determinant of the integral equation 
inverting the nonlocal $\bar\partial$-dressing problem with the $\tau$-function.
Finally, we establish equivalence between the Desargues maps and quadrilateral
lattices provided we take into consideration also their Laplace transforms. 
\end{abstract}
\maketitle

\section{Introduction}
Perhaps the most widely studied 
integrable discrete system of equations is the Hirota--Miwa system
\begin{equation} \label{eq:H-M}
\tau_{(i)}\tau_{(jk)} - \tau_{(j)}\tau_{(ik)} + \tau_{(k)}\tau_{(ij)} =0,
\qquad 1\leq i< j <k \leq N,
\end{equation}
which is the compatibility condition of the linear equations
(the adjoint of the introduced in~\cite{DJM-II})
\begin{equation} \label{eq:lin-dKP-comm}
\bphi_{(i)} - \bphi_{(j)} = \frac{\tau \tau_{(ij)}}{\tau_{(i)}\tau_{(j)}} \bphi ,
  \qquad 1\leq i < j \leq N.
\end{equation}
Here and in all the paper we use the convention that
for any function $f$ defined on multidimensional integer lattice $\ZZ^N$ by
$f_{(\pm i)}$ we denote its shift in the $i$ (positive or negative)
direction of the lattice, i.e.,
$f_{(\pm i)}(n_1,\dots,n_i,\dots,n_N) = f(n_1,\dots,n_i\pm 1,\dots,n_N)$. 
Whenever it does not lead to misunderstanding, when speaking on the 
image $f(n)$ of a point $n\in\ZZ^N$, we skip the argument.

In the basic case $N=3$, when the system \eqref{eq:H-M} reduces to a single
equation, it was discovered, up to a change of independent variables, by 
Hirota \cite{Hirota}, who called it 
the discrete analogue of the two dimensional Toda lattice, as a culmination 
of his studies on the bilinear form of 
nonlinear integrable equations; see \cite{Zabrodin} for a
review of various forms of the equation and of
its reductions.
General feature of Hirota's
equation was uncovered by Miwa \cite{Miwa} who found a remarkable transformation
which connects the equation to the Kadomtsev--Petviashvili (KP) 
hierarchy~\cite{DKJM}. The Hirota--Miwa
equation/system can be encountered in various
branches of theoretical physics \cite{Saito,KNS} and mathematics
\cite{Shiota,KWZ,Octahedron}. In the literature there are known also
non-commutative versions \cite{FWN,FWN-Capel,Nimmo-NCKP} of the Hirota--Miwa
system. 

During last years there was some activity in providing geometrical
interpretation for integrable discrete systems. The idea was to transfer 
to a
discrete level the well known connection between geometry and integrable
differential equations, see classical monographs
\cite{DarbouxIV,Darboux-OS,Bianchi,Eisenhart-TS,Tzitzeica,Finikov} written in
the pre-solitonic period, and more recent works
\cite{Sym,RogersSchief,GuHuZhou}. Almost after the first works in this direction,
which included the
discrete pseudospherical surfaces \cite{BP1}, evolutions of discrete curves
\cite{DS-AL}, and discrete isothermic surfaces \cite{BP2}, in 
\cite{DCN} there was given a geometric interpretation of the $N=3$ dimensional
Hirota--Miwa equation in its two dimensional Toda lattice form. The
basic geometric object in \cite{DCN} was the Laplace sequence
of two dimensional lattices made of planar quadrilaterals; see also
Section \ref{sec:D-QL} for more details. Such lattices were
introduced much earlier \cite{Sauer2,Sauer} as discrete analogs of conjugate
nets on a surface. 

\begin{figure}
\begin{center}
\includegraphics{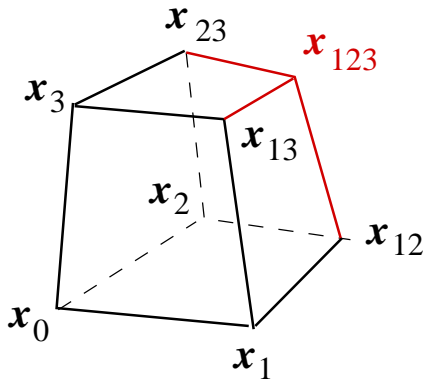}
\end{center}
\caption{The geometric integrability scheme}
\label{fig:TiTjTkx}
\end{figure}

Soon after \cite{DCN} the multidimensional lattices of planar
quadrilaterals, called also quadrilateral lattices for short, were
considered in \cite{MQL}. In particular, it was shown there 
that such
lattices are described by solutions of the discrete
Darboux system \cite{BoKo}. The initial boundary value problem for
multidimensional quadrilateral lattice is based on the following simple
geometric statement (see Figure~\ref{fig:TiTjTkx}).

\emph{Consider points $x_0$, $x_1$, $x_2$ and $x_3$ in general position in $\PP^M$,
$M\geq 3$. On
the plane $\langle x_0, x_i, x_j \rangle$, $1\leq i < j \leq 3$ choose a point
$x_{ij}$ not on the lines  $\langle x_0, x_i \rangle$, $\langle x_0,x_j
\rangle$ and $\langle x_i, x_j \rangle$. Then there exists the
unique point $x_{123}$
which belongs simultaneously to the three planes 
$\langle x_3, x_{13}, x_{23} \rangle$,
$\langle x_2, x_{12}, x_{23} \rangle$ and
$\langle x_1, x_{12}, x_{13} \rangle$.
}

This construction scheme is multidimensionally compatible 
\cite{MQL} and provides  initial boundary value problem for the corresponding 
discrete Darboux equations in terms of $K(K-1)$ functions of two discrete 
variables (implying also multidimensional compatibility of the system). As it 
was shown in \cite{TQL}, the
Darboux transformations of the quadrilateral lattice (thus also the
corresponding B\"{a}cklund transformations of the discrete Darboux equations),
being discrete symmetries of the quadrilateral lattice Darboux system,
can be considered as recursive augmentation of the number of independent
variables. Moreover the Bianchi permutability principle of superposition of the
transformations, which is often considered as synonymous to integrability, 
is a consequence of that simple geometric scheme. Therefore the "compatibility 
of the construction for arbitrary dimension of the lattice" \cite{MQL}
provides its own commuting symmetries. 

We remark that the discrete Darboux
equations have been constructed in \cite{BoKo} as the most general discrete 
system integrable by the nonlocal
$\bar\partial$--dressing method. Moreover the differential Darboux equations
\cite{Darboux-OS} give in a special limit \cite{BogdanovManakov} the KP
hierarchy of equations (in this context the possibility of considering
arbitrary number of independent
variables is crucial). This places the quadrilateral lattice (Darboux)
equations in a distinguished position among all integrable discrete systems; see
also works \cite{BoKo-N-KP,DMMMS,DMM} on the relation of Darboux equations,
conjugate nets and quadrilateral
lattices, and the multicomponent KP hierarchy \cite{DKJM}, which is often
considered as the master integrable system. 

Having recognized the
role of the quadrilateral lattice as the master geometric object of the
integrability theory the remaining task is to study integrable reductions of the
lattice. If a geometric
constraint, imposed on the initial points, 
propagates during the
construction, the corresponding reduction of the discrete Darboux equations is
called geometrically integrable. Notice that such consistency of the geometric 
integrability scheme with the  reduction
in conjunction with the multidimensional
compatibility of the quadrilateral lattice implies the multidimensional
compatibility of the reduction. This point of view was used in 
\cite{CDS,q-red,DS-sym,BQL,CQL}, see also recent
review \cite{DS-EMP}, to select integrable reductions of the
quadrilateral lattice and to find
the corresponding reductions of the discrete Darboux
equations. 

For example, the integrability of quadrilateral
lattices with elementary quadrilaterals inscribed in circles, introduced in
\cite{Bobenko-O} as discrete analog of orthogonal coordinate systems, was first
proved in this way in \cite{CDS}. The integrability of the circular lattice was
then confirmed by the nonlocal
$\bar\partial$-dressing method \cite{DMS}, by construction of the corresponding
Darboux-type transformation \cite{KoSchief2} which satisfies
\cite{LiuManas-SR,q-red} the permutability property, by construction of such
lattices using the Miwa transformation from the multicomponent BKP hierarchy
\cite{DMM}, and by application of the algebro-geometric techniques \cite{AKV}.
Remarkably, as  described
in \cite{BaMaSe}, there exists a quantization procedure for circular lattices, 
which leads to solutions of the tetrahedron equation (the
three dimensional analog of the Yang--Baxter equation).

In \cite{ABS} (see also \cite{FWN-cons}), the notion of 
multidimensional consistency as a tool to detect integrable equations has been
given in the following
form: a $d$--dimensional discrete equation possesses the
\textbf{consistency} property, if it may be imposed in a consistent way on all
$d$--dimensional sublattices of a $(d+1)$--dimensional lattice. In the
terminology of \cite{ABS} the linear problem of the quadrilateral lattice
is three--dimensionally consistent, while the
discrete Darboux system is four--dimensionally consistent.\footnote{Notice a
terminological confusion: following
\cite{MQL} we speak on "compatibility of the construction
for arbitrary dimension of the lattice", while 
the word "consistency" is used in the context of integrable
reductions of the quadrilateral lattice (see, for example \cite{TQL}, where we 
say "geometric integrability scheme is consistent with the circular
reduction").}

It turns out that the
geometric notion of integrability of reductions of the quadrilateral lattice
very often associates them
with classical theorems \cite{Coxeter-IG} of incidence geometry. For example, 
integrability of the
circular lattice is a consequence of the Miquel theorem \cite{Miquel}. This
observation makes the relation between integrability of the discrete systems
and geometry even more
profound then the corresponding relation on the level of
differential equations. Integrable reductions of the quadrilateral 
lattice come from two sources.
The first are inner (i.e. invariant
with respect to the full group of projective transformations of the ambient
space) symmetries of the lattice. The second type of reductions arises
from the postulated existence of
additional structures (e.g., distinguished quadrics, hyperplanes) in the ambient
space and mimics the Cayley--Klein approach to
subgeometries of the projective geometry, which was the starting point of the
famous Erlangen program. Such approach to possible classification of 
integrable
discrete systems was formulated in \cite{q-red,W-cong}, see
also~\cite{BobSur-Proc,BobSur}. 

Apart from the geometric interpretation of the three dimensional
Hirota--Miwa equation in its two dimensional Toda lattice form,
there is known in the literature~\cite{KoSchief-Men} 
an interpretation of its Schwarzian form, the so called Menelaus 
lattice. It is related to
the adjoint linear problem of \eqref{eq:lin-dKP-comm} for a map 
$\bphi^*:\ZZ^3\to\RR^M$ in the affine gauge, and gives the so called discrete
Schwarzian KP equation, which is related to the Hirota--Miwa equation by
a nonlocal transformation~\cite{KoSchief-Men,Schief-JNMP,KingSchief1}; 
see also Section~\ref{sec:alg}.
\begin{figure}
\begin{center}
\includegraphics[width=13cm]{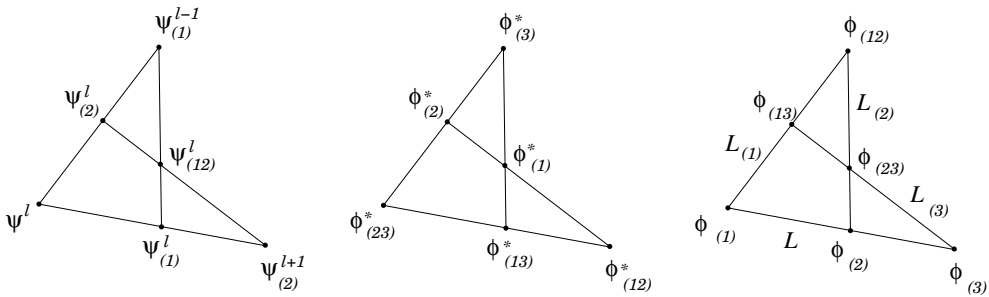}
\end{center}
\caption{Geometric equivalence of the Laplace sequence  $\psi^\ell$
of two dimensional quadrilateral lattices, the Menelaus lattice $\phi^*$, and
three dimensional Desargues lattice~$\phi$.}
\label{fig:T-M-D}
\end{figure}

An important observation~\cite{Bobenko-talk}, which was one of motivations 
of the present research, associates  
the four dimensional consistency  
of the discrete Schwarzian KP equation
with the Desargues configuration\footnote{In \cite{Bobenko-talk} we find
reference of this fact to the work of A.~D.~King and W.~K.~Schief \cite{KingSchief1},
where however this result is not mentioned. I have learned \cite{Schief-private}
that this important observation is due to W.~K.~Schief.}; see Section~\ref{sec:geom}.
Another fact, which was the starting point of the paper, is that 
there is no essential difference between the space 
of the algebro-geometric solutions of the 
Hirota--Miwa equations \cite{Krichever-4p,KWZ} and 
the quadrilateral lattice Darboux system \cite{AKV}, provided one
takes their Laplace transforms \cite{TQL} into consideration 
\cite{AD-Chicago}. 

In the paper we study the maps $\phi:\ZZ^N\to\PP^M$ defined by the most simple
nontrivial linear condition stating that for any pair of indices $i\ne j$ 
the points $\phi$, $\phi_{(i)}$ and $\phi_{(j)}$ are collinear. This is a
natural geometric counterpart of the linear problem~\eqref{eq:lin-dKP-comm}.
We show in a synthetic geometry way that
the multidimensional compatibility of the map
follows from the Desargues theorem and its
higher-dimensional analogs. 

Then, in
Section~\ref{sec:alg} we draw algebraic consequences of the geometric
definition of the
Desargues maps. As the algebraic significance of the 
Desargues theorem suggest \cite{BeukenhoutCameron-H}, 
we consider projective spaces over
division rings, what leads to the non-Abelian Hirota--Miwa system
\cite{Nimmo-NCKP}. We discuss also various gauge-equivalent forms of the 
equation in the non-commutative setting. 

It can be seen both from simple geometric and
algebraic considerations  that the Desargues maps can be called also 
multidimensional adjoint
Menelaus maps; see Figure~\ref{fig:T-M-D}. We prefer however
to call it in a way which reflects the projective geometric
character of the lattice and captures simultaneously its integrability
properties.
 
In Section~\ref{sec:D-bar} we apply the nonlocal $\bar\partial$-dressing method
\cite{AYF,ZakhMan,Konopelchenko} to find large classes of solutions to the
Hirota--Miwa system over the field $\CC$ of complex numbers. 
In particular we show, as one may expect from works
\cite{Palmer,PoppeSattinger,SegalWilson,AD-tauQL}, that the $\tau$-function of the
Hirota--Miwa system can be identified with the Fredholm determinant 
of the integral equation inverting the nonlocal $\bar\partial$-dressing problem.
We find that also on the level of the nonlocal $\bar\partial$-dressing method
the solution space of the Hirota--Miwa system is the same like in the case
of quadrilateral lattice Darboux system~\cite{BoKo} provided one 
takes~\cite{TQL} also the Laplace
transformations of the lattice into consideration.

The three point condition in definition of the Desargues map
can be considered as a serious
degeneration of the quadrilateral lattice map four point
condition. Such approach was presented for the three point linear problem
of the Menelaus lattice for example in~\cite{KoSchief-Men}.  
In Section~\ref{sec:D-QL} we
show however that the quadrilateral lattice theory and the Desargues 
lattice theory are equivalent. 

After submitting the first version of the manuscript I have learnt on 
related recent works of W.~K.~Schief \cite{Schief-talk} and V.~E.~Adler
\cite{Adler-tangential}.

\section{Geometry of the Desargues maps}
\label{sec:geom}
In this Section we study in detail geometric properties of the Desargues maps.
After collecting basic facts on the Desargues configuration we state some
genericity assumptions concerning the maps. Then we study multidimensional
compatibility of the Desargues maps. Here we understand this notion as a
possibility of recursive augmentation of the number of independent variables
preserving the geometric condition that characterizes the maps. This point of
view mimics successive application of the Darboux transformations
or the recursion operator \cite{Olver-rec}. We
postpone discussion of the initial boundary value problem for the Desargues
maps to Section~\ref{sec:D-QL} after we show their relation to the 
quadrilateral lattices.
\begin{figure}
\begin{center}
\includegraphics[width=7cm]{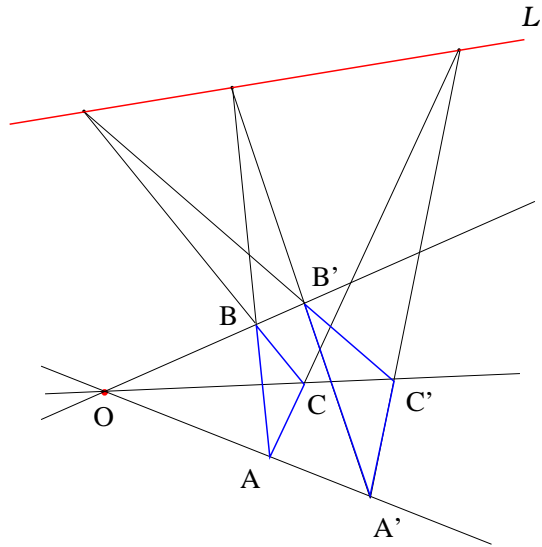}
\end{center}
\caption{The triangles $\triangle ABC$ and
$\triangle A^\prime B^\prime C^\prime $ are perspective from the point $O$ if
and only if they are perspective from the line $L$.}
\label{fig:Desargues}
\end{figure}

\subsection{The Desargues configuration}
Among all incidence theorems in projective geometry the Desargues theorem (see
Figure~\ref{fig:Desargues}) plays a very distinguished
role~\cite{Coxeter-IG,BeukenhoutCameron-H}.
It holds in projective spaces of dimension more then two, and is an important
element in proving the possibility of introduction of homogeneous coordinates
taking values in a division ring; in order to introduce such coordinates on 
projective planes one should add it as an axiom.

The ten lines involved and the ten points involved are so arranged that each 
of the ten
lines passes through three of the ten points, and each of the ten points lies 
on three
of the ten lines.
Under the standard duality of plane projective geometry (where points 
correspond to
lines and collinearity of points corresponds to concurrency of lines), 
the Desargues
configuration is self-dual: axial perspectivity is translated into central
perspectivity and vice versa.  

At first sight it seems that the Desargues
configuration has less symmetry than it really has. However, any of the ten 
points may 
be chosen
to be the center of perspectivity, and that choice determines which six points 
will be
vertices of triangles and which line will be the axis of perspectivity. 
The Desargues
configuration has symmetry group $S_5$ of order $120$. It
can be constructed from a $5$ point set, preserving the action of the 
symmetric group,
by letting the points and lines of the Desargues configuration correspond to 
$2$ and
$3$ element subsets of the $5$ points, with incidence corresponding to 
containment.

\begin{Rem}
In the above interpretation of the symmetry group of the Desargues
configuration, the $4$ element subsets give rise to $5$ complete
quadrilaterals described by the Menelaus theorem, as used in
\cite{Bobenko-talk} in connection with 
the four dimensional consistency of the discrete Schwarzian KP
equation.  
\end{Rem}

\subsection{The Desargues maps}

In the paper we study the following maps, the connection of which with the
Desargues theorem 
is essential in showing their multidimensional
compatibility.
\begin{Def} \label{def:Des-map}
By \emph{Desargues map} we mean a map $\phi:\ZZ^N\to\PP^M$ of
multidimensional integer lattice in Desarguesian
projective space of dimension $M\geq 2$,
such that for any pair of indices $i\ne j$ the points $\phi$, $\phi_{(i)}$ and
$\phi_{(j)}$ are collinear.
\end{Def}
\begin{Rem}
Notice that we consider $\ZZ^N$ as a directed graph.
\end{Rem}
\begin{Rem}
The image of a Desargues map can be called a \emph{Desargues lattice}. 
However we would like to stress that we do not use this notion in the sense of 
the lattice theory as described in~\cite{Birkhoff,Jonsson}.
\end{Rem}
Let us discuss various genericity assumptions of the map.
Consider an $N$-dimensional, $N>0$, hypercube graph
with a distinguished vertex labeled by ${\emptyset}$, its first order neighbours
labeled by ${\{i\}}$, $i=1,\dots , N$, and other vertices labeled 
as follows: the fourth vertex of a quadrilateral with three other vertices 
${I}$, ${I\cup\{i\}}$, ${I\cup\{j\}}$, $i,j\not\in I$, $i\not= j$ is
${I\cup\{i,j\}}$.
\begin{Def}
\emph{A Desargues $N$-hypercube} consists of labelled vertices $\phi_I$ of 
an $N$ dimensional hypercube in projective space 
$\PP^M$, $M\geq 2$, such that for arbitrary multiindex
$I\varsubsetneq \{ 1,2, \dots , N\}$ there exists a line $L_I$ incident
with $\phi_I$ and with all the points $\phi_{I\cup\{i\}}$, $i\notin I$. 
A Desargues
$N$-hypercube is called \emph{non-degenerate} if all its vertices are distinct.
A non-degenerate Desargues
$N$-hypercube is called \emph{weakly generic} if all the lines $L_I$ 
are distinct.
\end{Def}
Given two multiindices $I_1,I_2$, with $I_1\subset I_2$, 
the points $\phi_J$, $I_1 \subset J \subset I_2$, 
of a weakly generic Desargues $N$-hypercube form weakly generic Desargues
$(|I_2|-|I_1|)$-hypercube. The space $\pi_{I_1,I_2}$ spanned
by the points $\phi_J$, $I_1 \subset J \subset I_2$, has dimension
$(|I_2|-|I_1|)$ at most. For example, $L_I = \pi_{I,I\cup \{i\}}$ for all
$i\notin I$. We write also $\pi_{I}=\pi_{\emptyset,I}$.
\begin{Def}
A Desargues $N$-hypercube is called \emph{generic} if $\dim \pi_{I_1,I_2}
= |I_2| - |I_1|$ for all $I_1 \subset I_2$.
\end{Def}
\begin{Rem}
Notice that suitable projections of a generic Desargues hypercubes can produce
weakly generic Desargues hypercubes.
\end{Rem}
\begin{Def}
A  Desargues map $\phi:\ZZ^N\to\PP^M$
is called \emph{(weakly) generic} if the corresponding 
Desargues lattice consists of (weakly) generic Desargues $N$-hypercubes under 
identification $\phi_I$ with $\phi_{(I)}$ for a fixed point $\phi$ of the
lattice.
\end{Def} 

Notice that any weakly generic Desargues map
$\phi:\ZZ^N\to\PP^M$ induces a map $L:\ZZ^N\to\GG^M_1$ into the Grassmann 
space of lines in $\PP^M$, where $L$ is the line coincident with the point 
$\phi$ and all the neighbouring $\phi_{(i)}$, $i=1,\dots, N$. Such maps are
characterized by the following two properties.\\
(i) Any two neighbouring lines $L$ and $L_{(i)}$ intersect.\\
(ii) The intersection points $L\cap L_{(-i)}$ coincide for all
$1\leq i\leq N$.

The maps $L:\ZZ^N\to\GG^M_1$ satisfying the first condition 
only, play an important
role in the theory of Darboux transformations of the quadrilateral
lattice~\cite{TQL} and are called \emph{line congruences}. It is natural to call
the line congruences satisfying also the second condition the
\emph{Desargues congruences}.
Then the points of the Desargues lattice can be recovered by 
$\phi= L\cap L_{(-i)}$, $i=1,\dots , N$.

\subsection{Multidimensional compatibility of Desargues maps}

Given point $\phi$ and its two nearest 
(in positive directions)
neighbours  $\phi_{(i)}$ and $\phi_{(j)}$. By definition there exists a line $L$
incident with the three points. 
Assuming the Desargues map is weakly generic, the point 
$\phi_{(ij)}$ can be an arbitrary point not on the line $L$. Such a choice
determines the lines $L_{(i)}$ and $L_{(j)}$.

\subsubsection{Three dimensional compatibility and the Veblen--Young axiom}

Consider a point $\phi_{(k)}\in L$, $k\ne i,j$. On the line $L_{(i)}$ choose 
a point $\phi_{(ik)}$ 
distinct from $\phi_{(i)}$ and $\phi_{(ij)}$, thus determining the line
$L_{(k)}$. Then three dimensional compatibility of the Desargues map, i.e. the
existence of the intersection point $\phi_{(jk)}$ of lines 
$L_{(j)}$ and $L_{(k)}$, is
equivalent to the Veblen-Young axiom of the synthetic projective geometry, 
which in the current notation states (compare
Figure~\ref{fig:T-M-D}).

\emph{Given four distinct points $\phi_{(j)}$, $\phi_{(k)}$,
$\phi_{(ij)}$, $\phi_{(ik)}$; if the lines $L_{\phi_{(j)}\phi_{(k)}}= L$ and
$L_{\phi_{(ij)}\phi_{(ik)}}=L_{(i)}$ intersect, then the lines 
$L_{\phi_{(j)}\phi_{(ij)}}= L_{(j)}$ and $L_{\phi_{(k)}\phi_{(ik)}} =L_{(k)}$ 
intersect as well.} 

There is no condition for the point $\phi_{(ijk)}$, apart from
weak genericity assumption, which means that it should
not be placed on the lines $L$, $L_{(i)}$, $L_{(j)}$, $L_{(k)}$. 

\subsubsection{Four dimensional compatibility and the Desargues theorem}

Add the next point
$\phi_{(\ell)}$ on the line $L$, $\ell \ne i,j,k$, and the point 
$\phi_{(i\ell)}\in L_{(i)}$. The corresponding line $L_{(\ell)}$, 
incident with $\phi_{(\ell)}$
and $\phi_{(i\ell)}$, intersects (by Veblen--Young) the lines $L_{(j)}$,
$L_{(k)}$ in the points $\phi_{(j\ell)}$ and $\phi_{(k\ell)}$, correspondingly. 
The problem is to find the four points 
$\phi_{(ijk)}$, $\phi_{(ij\ell)}$,
$\phi_{(ik\ell)}$ and $\phi_{(jk\ell)}$ which satisfy the Desargues map 
condition.
\begin{figure}
\begin{center}
\includegraphics[width=7cm]{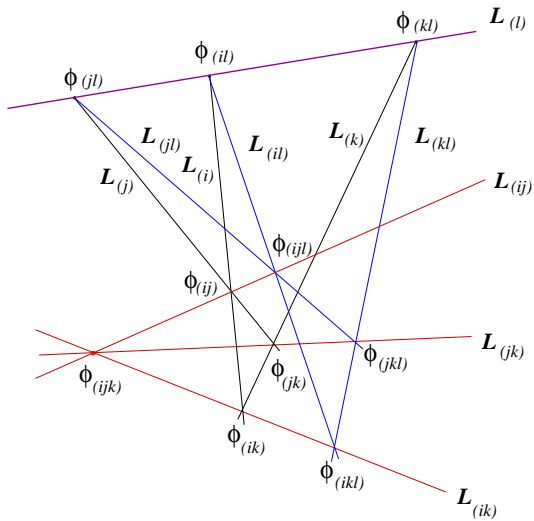}
\end{center}
\caption{The four dimensional compatibility of the Desargues map}
\label{fig:Desargues-phi}
\end{figure}

Choose a point $\phi_{(ij\ell)}$, not on the lines $L$, $L_{(i)}$, $L_{(j)}$,
$L_{(k)}$, and define therefore the lines $L_{(ij)}$, $L_{(i\ell)}$ and 
$L_{(j\ell)}$. On the line $L_{(i\ell)}$ mark a point $\phi_{(ik\ell)}$, thus
defining the lines $L_{(ik)}$ and $L_{(k\ell)}$. 
The lines $L_{(j\ell)}$ and $L_{(k\ell)}$ intersect (by Veblen--Young) 
in the point
$\phi_{(jk\ell)}$, which gives the line $L_{(jk)}$. 
We have constructed
two triangles in perspective from the line $L_{(\ell)}$: the first with
vertices $\phi_{(ij)}$, $\phi_{(ik)}$, $\phi_{(jk)}$, and the second with
vertices $\phi_{(ij\ell)}$, $\phi_{(ik\ell)}$, $\phi_{(jk\ell)}$.
By the Desargues theorem the three lines  $L_{(ij)}$, $L_{(ik)}$ and 
$L_{(jk)}$ intersect in one point, which is by construction $\phi_{(ijk)}$.

\begin{Rem}
Notice that in the generic case when the points $\phi$, $\phi_{(i)}$,
$\phi_{(ij)}$ and $\phi_{(ijk)}$ generate the space $\pi_{(ijk)}$ of 
dimension three, then
all the points whose shifts contain the index $\ell$ are obtained as
intersections of the lines of the "$ijk$ configuration" with the plane generated
by the points $\phi_{(\ell)}$, $\phi_{(i\ell)}$ and $\phi_{(ij\ell)}$. Moreover,
to keep the configuration generic we add the point $\phi_{(ijk\ell)}$
(which is not specified by
the previous construction) outside the space $\pi_{(ijk)}$ thus generating the
four dimensional space $\pi_{(ijk\ell)}$.
\end{Rem}

\subsubsection{The multidimensional compatibility for arbitrary $N$}
The multidimensional compatibility of the Desargues map is equivalent to existence
of a Desargues $N$-hypercube for arbitrary $N$, provided appropriate initial
data have been prescribed. 
The following proposition allows to construct generic Desargues
$(N+1)$-hypercubes from generic Desargues
$N$-hypercubes in spaces of the dimension large enough. 
It is an analogue of the well known, mentioned in the Remark above, 
three dimensional proof of the Desargues theorem.
By suitable projections one can produce therefore
weakly generic Desargues hypercubes.
\begin{Prop} \label{prop:N->N+1}
Given generic Desargues $N$-hypercube in $\PP^M$, where $N<M$.  
On the $N$ lines $L_\emptyset$, $L_{\{1\}}$, $L_{\{1,2\}}$,  
$L_{\{1,2 , \dots , N-1\}}$, chose $N$ points $\phi_{\{N+1\}}$,
$\phi_{\{1,N+1\}}$, $\phi_{\{1,2,N+1\}}$, \dots 
$\phi_{\{1,2,\dots , N-1, N+1\}}$ in generic position, correspondingly, 
in such a way that the $N-1$ dimensional subspace $U$ of 
$\pi_{\{ 1,2, \dots , N\}}$
\begin{equation*}
U=\langle  \phi_{\{N+1\}}, \phi_{\{1,N+1\}}, \phi_{\{1,2,N+1\}}, \dots 
\phi_{\{1,2,\dots , N-1, N+1\}} \rangle
\end{equation*}  
is not incident with any vertex of the hypercube. 
Then the unique intersection points $\phi_{\{I,N+1\}}=U\cap L_I$ of the
hyperplane with the lines of the $N$-hypercube, and the points of the initial
hypercube supplemented by a point
$\phi_{\{ 1,2, \dots , N, N+1 \}} \notin \pi_{\{ 1,2, \dots , N\}}$ 
give a generic Desargues $(N+1)$-hypercube.
\end{Prop}
\begin{proof}
By the assumption of the Proposition the lines $L_I$,
$I\varsubsetneq \{ 1,2, \dots , N\}$, are not contained in the 
hyperplane $U$, thus all the points $\phi_{I\cup\{N+1\}}$ are well defined.
Having then all the vertices of the $(N+1)$-hypercube we will check that it
satisfies the desired properties. 

Given multiindex
$I\varsubsetneq \{ 1,2, \dots , N, N+1\}$ there are two possibilities.\\
(i) $N+1 \notin I$. When $|I|=N$ then 
$I=\{ 1,2, \dots , N\}$ and define $L_I$
as the unique line incident with $\phi_{\{ 1,2, \dots , N\}}$ and 
$\phi_{\{ 1,2, \dots , N, N+1\}}$.
If $|I| < N$ then take as the line $L_I$ the line of the $N$-hypercube,
and $\phi_{I\cup\{N+1\}} \in L_I$ by construction. \\
(ii) $N+1 \in I$. There exists $i\in\{ 1,2, \dots , N\}$, $i\notin I$.
Set $L_I$ as the unique line incident with $\phi_I$ and 
$\phi_{I\cup \{ i\}}$. To conclude the proof of the Desargues property
we will show that $L_I$ is independent of a particular choice of such
an index $i$. When $|I|=N$ then there is
nothing to prove because there is only one index $i\notin I$. 
If $1\leq |I| < N$ set $J = I\setminus \{N+1\}$, there exists 
$j\in \{ 1,2, \dots , N\}$, $i\not=j$, and $j \notin J$. Consider the plane
$\pi_{J,J\cup\{i,j\}}$ which contains the three lines $L_J$, $L_{J\cup\{i\}}$ and
$L_{J\cup\{j\}}$, and is not contained in $U$. Then 
$L_I =\pi_{J,J\cup\{i,j\}}\cap U$ which shows that also
$\phi_{J\cup\{j,N+1\}}=\phi_{I\cup\{j\}} \in L_I$, thus also the index $j$ can
be used to define $L_I$.

Finally, to prove genericity of the Desargues $(N+1)$-hypercube notice that 
for all $I_1 \subset I_2 \subset \{ 1,2, \dots , N\}$ the intersection
$\pi_{I_1,I_2} \cap L_{I_2}$ is the point $\phi_{I_2}$ only. This implies
that $\dim \pi_{I_1\cup\{N+1\},I_2\cup\{N+1\}} = |I_2| - |I_1|$, and 
$\dim \pi_{I_1,I_2\cup\{N+1\}} = |I_2| - |I_1|+1$.
\end{proof}

\subsection{The adjoint Desargues maps}
To obtain the analogous geometric meaning 
of the adjoint of the linear problem of the
Hirota--Miwa system, define the adjoint Desargues maps (or multidimensional 
Menelaus
maps, if one restricts to affine part of $\PP^M$) as maps 
$\phi^*:\ZZ^N\to\PP^M$ 
such that for any pair of indices $i\ne j$ the points $\phi^*_{(i)}$, 
$\phi^*_{(j)}$ and $\phi^*_{(ij)}$ are collinear. This leads to the following
definition of an adjoint Desargues $N$-hypercube.
\begin{Def}
\emph{An adjoint Desargues $N$-hypercube} consists of labelled vertices of 
an $N$ dimensional hypercube in projective space 
$\PP^M$, $M\geq 2$, such that for arbitrary multiindex
$I\subset\{ 1,2, \dots , N\}$, $|I|>1$, and for any pair of distinct
indices $i,j\in I$ the vertex $\phi^*_I$ is incident with a line passing through 
$\phi^*_{I\setminus\{i\}}$ and $\phi^*_{I\setminus\{j\}}$. 
\end{Def}

One can notice that given
Desargues map $\phi:\ZZ^N\to\PP^M$, its superposition $\phi\circ\imath$
with the arrows inversion map $\imath:\ZZ^N\to\ZZ^N$, $\imath(n)=-n$, is an
adjoint Desargues map (and vice versa). Similarly, any Desargues $N$-hypercube
gives rise to the adjoint Desargues $N$-hypercube under identification 
$\phi^*_I = \phi_{\{1,2, \dots , N\}\setminus I}$.
The geometric theory of the adjoint 
Desargues map follows from that identification.

\section{Desargues maps and various non-commutative
discrete KP equations}
\label{sec:alg}
In this Section we study algebraic consequences of the geometric definition of
the Desargues map $\phi:\ZZ^N\to\PP^M$. Because to prove its multidimensional
compatibility we use only the Desargues theorem then the natural coordinates of
the projective space are elements of a division ring $\DD$. This leads to the
corresponding non-commutative nonlinear equations which we formulate first in
arbitrary gauge, i.e., keeping the freedom in rescaling the homogeneous
coordinates by a nonzero factor. Two basic specifications of the gauge are
discussed in the second part of this Section.
\subsection{The linear problem for the Desargues maps
and its compatibility conditions}
In the homogeneous coordinates $\bphi:\ZZ^N\to\DD^{M+1}_*$ (we consider right
vector spaces) the map can be
described in terms of the linear system
\begin{equation} \label{eq:lin-A}
\bphi + \bphi_{(i)} A_{ij} + \bphi_{(j)} A_{ji} = 0, \qquad i\ne j,
\end{equation}
where $A_{ij}:\ZZ^N\to \DD_*$ are certain non-vanishing functions.
\begin{Prop}
The compatibility of the linear system \eqref{eq:lin-A} is equivalent to
equations 
\begin{align} \label{eq:alg-cond}
& A_{ij}^{-1} A_{ik} + A_{kj}^{-1} A_{ki} = 1, \\
\label{eq:shift-cond}
&A_{ik(j)}A_{jk} = A_{jk(i)} A_{ik},
\end{align}
where the indices $i,j,k$ are distinct.
\end{Prop}

\begin{proof}
From the linear problem \eqref{eq:lin-A} for the pair $(i,k)$ find $\bphi_{(k)}$
in terms of $\bphi$ and $\bphi_{(i)}$. Similarly, find $\bphi_{(k)}$
from the equation for the pair $(j,k)$. Comparing the resulting relation between 
$\bphi$ and $\bphi_{(i)}$ and $\bphi_{(j)}$ with the linear problem
\eqref{eq:lin-A} for the pair $(i,j)$ gives, after some elementary algebra, the
first equation.

The compatibility of the linear problem \eqref{eq:lin-A} shifted in $k$
direction with two other similar equations involving three distinct indices
$i,j,k$ gives rise to a linear relation between $\bphi$, $\bphi_{(k)}$ and
$\bphi_{(ij)}$. Their linear independence implies the vanishing of the
corresponding coefficients
\begin{align}
A_{ik}^{-1} A_{kj(i)}^{-1} A_{ij(k)} + A_{jk}^{-1} A_{ki(j)}^{-1} A_{ji(k)} &
=0, \label{eq:comp-a}\\
1 + A_{ki}A_{ik}^{-1} A_{kj(i)}^{-1} A_{ij(k)} + 
A_{kj}A_{jk}^{-1} A_{ki(j)}^{-1} A_{ji(k)} &
=0, \label{eq:comp-b}\\
A_{jk(i)} A_{kj(i)}^{-1} A_{ij(k)} + A_{ik(j)} A_{ki(j)}^{-1} A_{ji(k)} &
=0. \label{eq:comp-c}
\end{align}
Equations \eqref{eq:comp-a} and \eqref{eq:comp-c} directly lead 
\eqref{eq:shift-cond}.

Using \eqref{eq:shift-cond} we can replace equations
\eqref{eq:comp-a} and \eqref{eq:comp-b} by
\begin{align}
A_{ik}^{-1}  A_{ij} A_{kj}^{-1} + A_{jk}^{-1}  A_{ji} A_{ki}^{-1}&
=0, \label{eq:comp-a-n}\\
1 + (A_{ki}-A_{kj}) A_{ik}^{-1}  A_{ij} A_{kj}^{-1} &
=0. \label{eq:comp-b-n}
\end{align}
We will show that equation \eqref{eq:comp-a-n} follows from the condition
\eqref{eq:alg-cond}. Indeed, starting from the identity
\begin{equation*}
A_{kj}( 1 - A_{kj}^{-1} A_{ki}) +A_{ki}( 1 - A_{ki}^{-1} A_{kj}) =0,
\end{equation*}
and using \eqref{eq:alg-cond} we get
\begin{equation*}
A_{kj} A_{ij}^{-1}A_{ik} + A_{ki}A_{ji}^{-1}A_{jk} =0,
\end{equation*}
equivalent to \eqref{eq:comp-a-n}.
Also equation \eqref{eq:comp-b-n} is a direct consequence of the condition
\eqref{eq:alg-cond}.
\end{proof}
\begin{Cor}
For any three distinct indices $i,j,k$ we can write down three distinct
equations of the form \eqref{eq:alg-cond}. However any
two of them imply the third one.
\end{Cor}
\begin{proof}
Equation \eqref{eq:alg-cond}, where the indices $i$ and $k$ enter symmetrically
is equivalent to
\begin{equation} \label{eq:alg-cond-mod}
A_{ki}^{-1} A_{kj} = \left( 1 - A_{ij}^{-1} A_{ik} \right)^{-1}.
\end{equation}
Adding equation \eqref{eq:alg-cond-mod} to the similar one with the indices
$k$ and $j$ exchanged one obtains equation \eqref{eq:alg-cond} with the indices
$i$ and $j$ exchanged
\begin{equation*}
A_{ji}^{-1} A_{jk}+ A_{ki}^{-1} A_{kj} = 
\left[ A_{ik}^{-1}\left( A_{ik} - A_{ij} \right) \right]^{-1} +
\left[ A_{ij}^{-1}\left( A_{ij} - A_{ik} \right) \right]^{-1} =
\left( A_{ij} - A_{ik} \right)^{-1} \left( A_{ij} - A_{ik} \right) =1.
\end{equation*}
\end{proof}
\begin{Cor}
Equations \eqref{eq:shift-cond} imply existence of the potentials
$\rho_i:\ZZ^N\to\DD_*$, unique up to functions of single variables $n_i$, 
such that
\begin{equation} \label{eq:A-rho}
\rho_{i(j)} = A_{ji}\rho_i, \qquad i \neq j.
\end{equation}
\end{Cor}

\subsection{Gauges}
We are still left with the possibility to apply the gauge transformation
\begin{equation}
\bphi = \tilde\bphi G,
\end{equation}
where $G:\ZZ^N\to\DD_*$ is an arbitrary non-vanishing function. Then
$\tilde\bphi$ satisfies the linear problem \eqref{eq:lin-A} with the
coefficients
\begin{equation}
\tilde{A}_{ij} = G_{(i)}A_{ij}G^{-1}.
\end{equation}
By fixing properties of $G$ one can arrive to relation between the coefficients
of the linear problem. We will discuss two gauges. First gauge, which because of
the geometric interpretation can be called the affine gauge, gives 
in the commutative case the discrete modified KP
system. The second gauge in the commutative case leads to the Hirota--Miwa 
system. 

\subsubsection{The modified discrete KP gauge}
\begin{Prop}
When the gauge function is a non-vanishing solution of the linear problem
\eqref{eq:lin-A} 
then the coefficients $\tilde{A}_{ij}$ are constrained by the
relation
\begin{equation} \label{eq:aff}
\tilde{A}_{ij} + \tilde{A}_{ji} = - 1  , \qquad i\neq j.
\end{equation}
\end{Prop}
\begin{Rem}
When as the solution of the linear problem is taken the last coordinate
$\phi^{M+1}$ of the homogeneous representation of the map then we obtain the
standard transition to the non-homogeneous coordinates.
\end{Rem}
\begin{Rem}
In the affine gauge the algebraic compatibility system
\eqref{eq:alg-cond} consists,
for any triple of distinct indices $i,j,k$,  of one independent equation.
\end{Rem}
It is convenient (we follow the reasoning presented in \cite{Schief-JNMP} in the
commutative case)
to rewrite the linear problem \eqref{eq:lin-A} subject to condition
\eqref{eq:aff} as
\begin{equation}
(\bphi_{(j)} - \bphi) = (\bphi_{(i)} - \bphi)B_{ij}, 
\end{equation}
where
\begin{equation}
B_{ij} = B_{ji}^{-1} = A_{ij}(1+A_{ij})^{-1}.
\end{equation}
Then the algebraic compatibility takes the form 
\begin{equation}
B_{ij} B_{jk} = B_{ik},
\end{equation}
which allows for introduction of a potential $\sigma:\ZZ^N\to\DD_*$ such that
\begin{equation}
B_{ij} = \sigma_{(i)}\sigma_{(j)}^{-1}.
\end{equation}
The second part of the
compatibility condition takes then the form of the non-commutative discrete
mKP system \cite{FWN-Capel}
\begin{equation}
(\sigma_{(i)}^{-1} - \sigma_{(j)}^{-1})\sigma_{(ij)} + 
(\sigma_{(j)}^{-1} - \sigma_{(k)}^{-1})\sigma_{(jk)} + 
(\sigma_{(k)}^{-1} - \sigma_{(i)}^{-1})\sigma_{(ki)} =0. 
\end{equation}

Finally, notice that due to the compatibility of the system
\begin{equation}
(\bphi_{(j)} - \bphi)\sigma_{(j)} = (\bphi_{(i)} - \bphi)\sigma_{(i)} ,
\end{equation}
each coordinate $\phi^k:\ZZ^N\to\DD$ of $\bphi$ satisfies the generalized
lattice spin system \cite{FWN-Capel}
\begin{equation}
(\phi_{(jk)} - \phi_{(k)})(\phi_{(jk)} - \phi_{(j)})^{-1}
(\phi_{(ij)} - \phi_{(j)})(\phi_{(ij)} - \phi_{(i)})^{-1}
(\phi_{(ik)} - \phi_{(i)})(\phi_{(ik)} - \phi_{(k)})^{-1} = 1, 
\end{equation}
called also
the non-commutative Schwarzian discrete KP system 
\cite{BoKo-N-KP,KoSchiefSDS-II}.

\subsubsection{The Hirota--Miwa system gauge}

In order to introduce the second gauge we need the following result.
\begin{Lem}
There exists non-vanishing function $G$ defined as a solution of the system
\begin{equation} \label{eq:gauge-H}
G_{(i)} A_{ij} = - G_{(j)} A_{ji} , \qquad i\neq j.
\end{equation}
\end{Lem}
\begin{proof}
The
algebraic compatibility of 
equations \eqref{eq:gauge-H} for three pairs of indices $i,j,k$,
has the form
\begin{equation} \label{eq:gauge-H-comp}
A_{jk}^{-1}A_{ji}A_{ij}^{-1} + A_{kj}^{-1}A_{ki}A_{ik}^{-1} = 0, 
\qquad i,j,k \quad \text{distinct}.
\end{equation}
It can be proved
by application of the algebraic compatibility condition \eqref{eq:alg-cond} 
starting from the identity
\begin{equation*}
(1 - A_{ik}^{-1}A_{ij})A_{ij}^{-1} +(1 - A_{ij}^{-1}A_{ik})A_{ik}^{-1} =0. 
\end{equation*}
\end{proof}
\begin{Prop}
The linear system \eqref{eq:lin-A} is gauge equivalent to the 
discrete linear problem of the non-Abelian Hirota--Miwa system
\cite{DJM-II,Nimmo-NCKP}
\begin{equation} \label{eq:lin-dKP}
\bphi_{(i)} - \bphi_{(j)} =  \bphi U_{ij},  \qquad i \ne j \leq N.
\end{equation}
\end{Prop} 
\begin{proof}
Take the gauge function $G$ as in Lemma above, which gives (we skip tildes)
\begin{equation}
A_{ij} = -A_{ji},
\end{equation} \label{eq:U-H-U}
and set $U_{ij} =  A_{ji}^{-1}$.
\end{proof}
In this gauge the compatibility conditions
\eqref{eq:alg-cond}-\eqref{eq:shift-cond} reduce
to the following systems for distinct triples $i,j,k$
\begin{align} \label{eq:alg-comp-U}
& U_{ij} + U_{jk} + U_{ki} = 0, \\
& \label{eq:U-rho} 
U_{kj}U_{ki(j)} = U_{ki} U_{kj(i)}.
\end{align}
This allows to introduce potentials $r_i:\ZZ^N\to\DD_*$ such that
\begin{equation} \label{eq:def-r}
r_{i(j)} = r_i U_{ij}, \qquad i\ne j;
\end{equation}
see \cite{Nimmo-NCKP} for further properties of the system. 
\begin{Rem}
In the commutative case the functions $r_i$ can 
be parametrized in terms of a 
single potential $\tau$
\begin{equation} \label{eq:r-tau}
r_i = (-1)^{\sum_{k<i}n_k}\frac{\tau_{(i)}}{\tau},
\end{equation}
which leads to the linear problem \eqref{eq:lin-dKP-comm}, while
and the algebraic compatibility 
\eqref{eq:alg-comp-U} gives the Hirota--Miwa system \eqref{eq:H-M}.
\end{Rem}

\section{Application of the nonlocal $\bar\partial$-dressing method}
\label{sec:D-bar}
In this Section the division ring $\DD$ is
replaced by the field $\CC$ of complex numbers. By application
of the nonlocal $\bar\partial$-dressing method
\cite{AYF,ZakhMan,Konopelchenko} we construct 
solutions of the Hirota--Miwa system
and the corresponding solutions of the linear
problem. 

Consider the following
integro-differential equation in the complex plane ${\mathbb C}$
\begin{equation} \label{eq:db-nonl}
\bar\partial \chi(\lambda) = \bar\partial \eta(\lambda) + 
\int_{\mathbb C}  R(\lambda,\lambda')\chi(\lambda') \: 
d\lambda'\wedge d\bar\lambda',
\end{equation}
where $R(\lambda,\lambda')$ is a given $\bar\partial$ datum, which decreases
quickly enough at $\infty$ in $\lambda$ and $\lambda^\prime$, and the 
function $\eta(\lambda)$, the normalization of the unknown $\chi(\lambda)$, is a
given rational function, which describes the polar behavior of $\chi(\lambda)$ in $\CC$
and its behavior at $\infty$: 
\begin{equation*}
\chi(\lambda) - \eta(\lambda) \to 0, \qquad 
\text{for} \quad |\lambda|\to\infty.
\end{equation*}
We remark that the dependence of $\chi(\lambda)$ and $R(\lambda,\lambda')$ on
$\bar\lambda$ and $\bar\lambda^\prime$ will be systematically omitted, for
notational convenience.

Due to the generalized Cauchy formula the nonlocal $\bar\partial$ 
problem~\eqref{eq:db-nonl} 
is equivalent to the following Fredholm integral equation of the 
second kind  
\begin{equation} \label{eq:Fredh-db}
\chi(\lambda) = \eta(\lambda) - \int_{\mathbb C} K(\lambda, \lambda') 
\chi(\lambda') d\lambda'\wedge d\bar\lambda',
\end{equation}
with the kernel
\begin{equation} \label{eq:Fredh-ker-db}
K(\lambda, \lambda') = \frac{1}{2\pi i} \int_{\mathbb C} 
\frac{R(\lambda'', \lambda') }{\lambda'' - \lambda}
d\lambda''\wedge d\bar\lambda''  .  
\end{equation}
Recall (see, for example \cite{Smithies}) that 
the Fredholm determinant $D$ is defined by the series
\begin{equation}  \label{eq:Fr-det}
D = 1 +
\sum_{m=1}^\infty \frac{1}{m!} \int_{\CC^m}
K \begin{pmatrix} \zeta_1 & \zeta_2 & \dots  & \zeta_m \\
\zeta_1 & \zeta_2 & \dots  & \zeta_m
\end{pmatrix} d\zeta_1 \wedge d\bar\zeta_1 \dots d\zeta_m \wedge d\bar\zeta_m \; ,
\end{equation}
where
\begin{equation*}  \label{eq:Fr-KK}
K \begin{pmatrix} \zeta_1 & \zeta_2 & \dots  & \zeta_m \\ 
\mu_1 & \mu_2 & \dots  & \mu_m
\end{pmatrix} = \det \begin{pmatrix} 
K(\zeta_i,\mu_j)
\end{pmatrix}_{1\leq i,j \leq m} .
\end{equation*}
For a non-vanishing Fredholm determinant
the solution of 
\eqref{eq:Fredh-db} can be written in the form
\begin{equation} \label{eq:F-sol}
\chi(\lambda) = \eta(\lambda) - \int_\CC \frac{D(\lambda,\lambda^\prime)}{D}
\eta(\lambda^\prime) d\lambda'\wedge d\bar\lambda',
\end{equation}
where the Fredholm minor is defined by the series
\begin{equation}  \label{eq:Fr-min}
D(\lambda,\lambda^\prime) = \sum_{m=0}^\infty \frac{1}{m!} 
\int_{\CC^m} K \begin{pmatrix} 
\lambda & \zeta_1 &  \dots  & \zeta_m \\ 
\lambda^\prime & \zeta_1 &  \dots  & \zeta_m
\end{pmatrix} d\zeta_1 \wedge d\bar\zeta_1 \dots d\zeta_m \wedge d\bar\zeta_m  \; .
\end{equation}

Let $\lambda_i\in\CC$, $i=1,\dots,N$ be distinct points of the
complex plane. Consider the following 
dependence of the kernel $R$ on the variables $n=(n_1,\dots ,n_N)\in\ZZ^N$
\begin{equation} \label{eq:evol-R-dis-1}
R_{(i)}(\lambda,\lambda^\prime; n) =  (\lambda- \lambda_i)^{-1} 
R(\lambda,\lambda^\prime) (\lambda^\prime - \lambda_i),
\end{equation}
or equivalently 
\begin{equation} \label{eq:evol-R-dis}
R(\lambda,\lambda^\prime; n) = 
E(\lambda; n)^{-1} 
R_0(\lambda,\lambda^\prime) E(\lambda^\prime; n),
\qquad
E(\lambda;n)= \prod_{i=1}^N (\lambda - \lambda_i)^{n_i} ,
\end{equation}
where $R_0(\lambda,\lambda^\prime)$ is independent of $n$. We assume that $R_0$
decreases at $\lambda_i$ and in poles of the normalization function $\eta$
fast enough such that $\chi - \eta$ is regular in these points
\cite{BogdanovManakov,BoKo}.
\begin{Rem}
In the paper we always assume that the kernel $R$ in the nonlocal 
$\bar\partial$
problem is such that the Fredholm equation \eqref{eq:Fredh-db}
is uniquely solvable. Then, by the Fredholm alternative, the homogeneous 
equation with $\eta=0$ has only the trivial solution. 
\end{Rem}
\begin{Rem}
The structure of the function $E(\lambda;n)$ mimics the analytic structure of
the Baker--Akhezer wave function used in \cite{KWZ} to solve the Hirota--Miwa
system by the algebro-geometric techniques, where the role of the Fredholm
alternative is played by the Riemann--Roch theorem.  
\end{Rem}

\begin{Lem} \label{lem:evol-K-m}
The evolution \eqref{eq:evol-K-dis} of the kernel of the Fredholm equation
implies the following evolution of the determinants in the series defining the 
Fredholm determinant $D$ 
\begin{equation*}
K_{(i)}\left( \begin{matrix} \zeta_1 & \dots  & \zeta_m \\
\zeta_1  & \dots  & \zeta_m \end{matrix} \biggr\rvert \; n \right)
= K\left( \begin{matrix} \zeta_1  & \dots  & \zeta_m \\
\zeta_1  & \dots  & \zeta_m \end{matrix} \biggr\rvert \; n \right) -
\sum_{j=1}^m 
K\left( \begin{matrix} \zeta_1 & \dots & \lambda_i   
& \dots & \zeta_m \\
\zeta_1 &  \dots & \zeta_j 
& \dots & \zeta_m \end{matrix} \biggr\rvert \; n \right).
\end{equation*}
\end{Lem}
\begin{proof}
The evolution \eqref{eq:evol-R-dis-1} of the kernel $R$ implies that the
kernel $K$ of the integral equation \eqref{eq:Fredh-db}
is subject to the equation 
\begin{equation} \label{eq:evol-K-dis} 
K_{(i)}(\lambda,\lambda^\prime; n)  = 
(\lambda - \lambda_i)^{-1} \left[ K(\lambda,\lambda^\prime; n) -
K(\lambda_i,\lambda^\prime;n) \right] (\lambda^\prime - \lambda_i), 
\end{equation}
the conclusion is reached by basic linear algebra.
\end{proof}
\begin{Prop} \label{prop:lin-DB}
Let $\chi(\lambda;n)$ be a solution of
the $\bar\partial$ problem \eqref{eq:db-nonl} with the canonical 
normalization $\eta = 1$ then the function $\psi(\lambda;n) = \chi(\lambda;n)
E(\lambda;n)$ satisfies the linear system \eqref{eq:lin-dKP} with the
potentials
\begin{equation} 
U_{ij}(n) = (\lambda_j - \lambda_i)
\frac{\chi_{(j)}(\lambda_i;n)}{\chi(\lambda_i;n)} .
\end{equation}
\end{Prop}
\begin{proof}
The combination
$(\lambda - \lambda_i) \chi_{(i)}(\lambda;n) - (\lambda - \lambda_j) 
\chi_{(j)}(\lambda;n)$
satisfies the Fredholm equation with constant (in $\lambda$) normalization thus
must be proportional to $\chi(\lambda;n)$. By evaluating of both sides in
$\lambda_i$ we find the coefficient of proportionality. Multiplication of both
sides by $E(\lambda;n)$ gives the statement.
\end{proof}
\begin{Cor}
The form of $U_{ij}$ given above implies that the potentials $r_i$, defined by
equation \eqref{eq:def-r}, read
\begin{equation} \label{eq:r-i-DB}
r_i(n) = \prod_{k\ne i}(\lambda_k - \lambda_i)^{n_k} \chi(\lambda_i;n).
\end{equation}  
\end{Cor}

\begin{Th}
Within the considered class of solutions of the Hirota--Miwa system the
$\tau$-function is given by
\begin{equation}
\tau(n) = (-1)^{\sum_{i < j}n_i n_j /2}\prod_{i\ne j} (\lambda_i -
\lambda_j)^{n_i n_j /2}   D(n) . 
\end{equation}
\end{Th}
\begin{proof}
Evaluation of equation \eqref{eq:F-sol} at $\lambda_i$ for $\chi(\lambda;n)$ 
like in Proposition~\ref{prop:lin-DB} gives
\begin{equation}
\chi(\lambda_i;n) = 1 - \int_\CC \frac{D(\lambda_i,\lambda^\prime;n)}{D(n)}
d\lambda'\wedge d\bar\lambda'.
\end{equation} 
From Lemma~\ref{lem:evol-K-m} we obtain that
\begin{equation}
D_{(i)}(n) = D(n) - \int_\CC D(\lambda_i,\lambda^\prime;n)
d\lambda'\wedge d\bar\lambda'.
\end{equation} 
Comparison of both equations shows that
\begin{equation}
D_{(i)}(n) = \chi(\lambda_i;n) D(n),
\end{equation}
which due to equations \eqref{eq:r-tau} and \eqref{eq:r-i-DB} gives the 
statement.
\end{proof}
The above result, provides the "determinant interpretation" of the
$\tau$-function within the class of solutions which can be obtained by 
application
of the nonlocal $\bar\partial$-dressing method. 

Recently, within the same
approach the $\tau$-function of the quadrilateral lattices has been studied
\cite{AD-tauQL}.
As it can be deduced from \cite{BoKo,TQL}, 
the structure of the $\bar\partial$ datum 
in the nonlocal $\bar\partial$-dressing method which leads to the quadrilateral
lattices and all the lattices generated by their Laplace
transforms is as follows. 
Let $\lambda_i^\pm\in\CC$, $i=1,\dots,K$ be 
pairs of distinct points of the complex plane, let $m=(m_1,  \dots,  m_K)\in
\ZZ^K$ be points of the $\ZZ^K$ integer lattice and let $\ell = (\ell_1, \dots ,
\ell_K)$, $\ell_1 + \ell_2 + \dots + \ell_K = 0$, be a point 
of the $A_{K-1}$
root lattice. The function $E(\lambda; (m,\ell))$ which should replace the
function $E(\lambda;n)$ in equation \eqref{eq:evol-R-dis} reads
\begin{equation} 
E(\lambda;(m, \ell))= \prod_{i=1}^K 
\frac{(\lambda -\lambda^-_i)^{m_i}}{(\lambda - \lambda^+_i)^{m_i + \ell_i}} .
\end{equation} 
The variable $m$ is the quadrilateral lattice discrete parameter,
while the Laplace transformation $\cL_{ij}$ is given by 
$\ell_i \mapsto \ell_i + 1$, $\ell_j \mapsto \ell_j - 1$.  
After the proper identification of $2K$ points  $\lambda_i^\pm$, $i=1,\dots ,K$, 
with  the points
$\lambda_1, \dots , \lambda_{2K-1}, \lambda_{2K}=\infty$, we obtain the 
change of variables
discussed in Section~\ref{sec:D-QL}.

\section{Desargues maps and quadrilateral lattices}
\label{sec:D-QL}
This Section is devoted to the study of the relation between Desargues maps and
quadrilateral lattices. We will show that the theory of quadrilateral
lattices can be embedded in the theory the Desargues maps, and for odd $N=2K-1$ this
embedding is one-to-one (the case of even $N=2K$ can be treated as 
dimensional reduction of $2K+1$). The relation described below generalizes 
the relation, known on the
 $\tau$-function level, between the Hirota--Miwa equation and its version 
 in the discrete two dimensional Toda lattice form
\cite{Zabrodin}. The relation between discrete two dimensional Toda lattice and
two dimensional quadrilateral lattice was the subject of \cite{DCN,Dol-Hir}.

Recall that  the condition of planarity of elementary quadrilaterals of
 $\psi:\ZZ^K \to \PP^M$
written in the non-homogeneous coordinates $\bpsi:\ZZ^K\to\DD^M$ gives 
the following linear problem
\begin{equation} \label{eq:lin-QL}
\bpsi_{(ij)} - \bpsi = (\bpsi_{(i)} - \bpsi )a_{ij} + 
(\bpsi_{(j)} - \bpsi )a_{ji}, \qquad i\neq j,
\end{equation}
where $a_{ij}:\ZZ^K\to\DD$ are certain functions which should satisfy the
corresponding compatibility condition (a version of the discrete Darboux system). 

\begin{figure}
\begin{center}
\includegraphics[width=9cm]{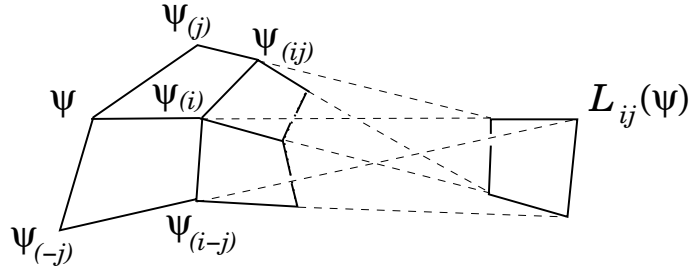}
\end{center}
\caption{The Laplace transform of the quadrilateral lattice}
\label{fig:Laplace}
\end{figure}

The Laplace transformation $\mathcal{L}_{ij}$
of $\psi$ is constructed \cite{DCN,TQL} via intersection of the tangent lines
$\langle \psi , \psi_{(i)} \rangle$ with their $j$-th negative neighbours
$\langle \psi_{(-j)} , \psi_{(i,-j)} \rangle$, see Figure~\ref{fig:Laplace}. 
In the non-homogeneous coordinates we have
\begin{equation} \label{eq:L-ij}
\mathcal{L}_{ij}(\bpsi) = \bpsi + (\bpsi_{(i)} - \bpsi) (1-a_{ji(-j)})^{-1}.
\end{equation}
The Laplace transforms of 
quadrilateral lattices are quadrilateral lattices again, and the following
relations hold \cite{TQL}
\begin{equation*}
\cL_{ij} \circ \cL_{ji}  = \text{id} \; ,\qquad
\cL_{jk} \circ \cL_{ij} = \cL_{ik}, \qquad
\cL_{ki} \circ \cL_{ij} = \cL_{kj}. 
\end{equation*}
They allow to parametrize the
quadrilateral lattices generated from one quadrilateral lattice via the Laplace
transformations by points of the root lattice of the type $A_{K-1}$ (see also
discussion in \cite{DMMMS}). This suggests to consider 
the Laplace transformation directions as new variables. In order to  place 
all variables on equal footing we change the variables as suggested in
Section~\ref{sec:D-bar}.

Consider, as  the following change of variables between $n\in\ZZ^{2K-1}$ 
integer lattice
and $(m,\ell)\in\ZZ^K\times Q(A_{K-1})$, where $Q(A_{K-1})=\{ \ell\in\ZZ^K \, | \,
\ell_1 + \ell_2 + \dots + \ell_K = 0 \}$ is the $A_{K-1}$  root lattice
\begin{equation*}
n_{2i-1} = m_i , \qquad n_{2i} = -m_i - \ell_i, \qquad i = 1, \dots , K,
\end{equation*}
here, for convenience,
we have defined also $n_{2K} = - n_1 -n_2 - \dots - n_{2K-1}$.

For fixed $\ell\in Q(A_{K-1})$ define the
map $\psi^\ell:\ZZ^K\to \PP^M$ given by
$\psi^\ell(m) = \phi(n)$, where the relation between $n$ and $m$ and $\ell$ is
given above. 
Then we have
\begin{equation}
\phi_{(\pm(2K-1))} = \psi^\ell_{(\pm K)},
\end{equation}
and for $i\ne K$
\begin{equation}
\phi_{(\pm(2i-1))} = \psi^{\ell \mp e_i \pm e_K}_{(\pm i)}, \qquad
\phi_{(\pm 2i)} = \psi^{\ell \mp e_i \pm e_K},
\end{equation}
where $e_i$ is the element of the canonical basis of $\RR^K$ having $1$ 
as $i$-th component and $0$'s elsewhere.
\begin{Prop}
The maps $\psi^\ell:\ZZ^K\to\PP^M$ are quadrilateral lattice maps. Moreover 
$\psi^{\ell + e_i - e_j}$ is the Laplace transform
$\mathcal{L}_{ij}(\psi^\ell)$ of $\psi^\ell$.
\end{Prop}
\begin{proof}
Assume that $i<j<K$.
The point $\phi_{(-2i,2j-1)} = \psi^{\ell + e_i - e_j}_{(j)}$ and the
points $\phi = \psi^\ell$, $\phi_{(2i-1,-2i)}= \psi^\ell_{(i)}$ belong to the
line containing (positive) neighbours of $\phi_{(-2i)}$.
Similarly, the same point
$\phi_{(-2i,2j-1)} = \psi^{\ell + e_i - e_j}_{(j)}$ and the
points $\phi_{(2j-1,-2j)} = \psi^\ell_{(j)}$, $\phi_{(2i-1,-2i,2j-1,-2j)}= 
\psi^\ell_{(ij)}$ belong to the
line containing (positive) neighbours of $\phi_{(-2i,2j-1,-2j)}$. This shows
that the lines $\langle\psi^\ell,\psi^\ell_{(i)}\rangle$ and
$\langle \psi^\ell_{(j)},\psi^\ell_{(ij)}\rangle$ intersect in 
$\psi^{\ell + e_i - e_j}_{(j)}$. Therefore the four
points $\psi^\ell$, $\psi^\ell_{(i)}$, $\psi^\ell_{(j)}$ 
and $\psi^\ell_{(ij)}$ are coplanar, and 
$\psi^{\ell + e_i - e_j} = \mathcal{L}_{ij}(\psi^\ell)$.

For  $j<i<K$  the reasoning is similar. The details of the case 
when one of the indices $i$ or $j$ is
equal to $K$ is left for the reader.
\end{proof}

Let us illustrate the above reasoning (still $i<j<K$)
in making simple calculation in the 
affine gauge \eqref{eq:aff}. Collinearity of $\psi^\ell$, $\psi^\ell_{(i)}$ and
$\psi^{\ell + e_i - e_j}_{(j)}$ gives
\begin{equation} \label{eq:Lij-psi}
(\bpsi^\ell_{(i)} - \bpsi^\ell) A_{2i-1,2i(-2i)} = 
(\bpsi^{\ell + e_i - e_j}_{(j)}- \bpsi^\ell)A_{2j-1,2i(-2i)} .
\end{equation}
Similarly, collinearity of $\psi^\ell_{(j)}$, $\psi^\ell_{(ij)}$ and
$\psi^{\ell + e_i - e_j}_{(j)}$ gives in the affine gauge
\begin{equation}
(\bpsi^{\ell}_{(ij)}-\bpsi^{\ell}_{(j)}) A_{2i-1,2i(-2i,2j-1,-2j)} = 
(\bpsi^{\ell + e_i - e_j}_{(j)} - \bpsi^\ell_{(j)})
A_{2j-1,2i(-2i,2j-1,-2j)}.
\end{equation}
Elimination of $\bpsi^{\ell + e_i - e_j}_{(j)}$ from the above equations
implies that $\bpsi^\ell$ satisfies equation \eqref{eq:lin-QL} 
with the coefficients
\begin{align}
a_{ij}^\ell & = A_{2i-1,2i(-2i)}A_{2j-1,2i(-2i)}^{-1} 
A_{2j-1,2i(-2i,2j-1,-2j)}A_{2i-1,2i(-2i,2j-1,-2j)}^{-1},\\
\label{eq:aji-l}
a_{ji}^\ell & = 1 - A_{2j-1,2i(-2i,2j-1,-2j)}A_{2i-1,2i(-2i,2j-1,-2j)}^{-1}.
\end{align}
Equation \eqref{eq:Lij-psi} gives
\begin{equation}
\bpsi^{\ell + e_i - e_j}_{(j)} = \bpsi^\ell +
(\bpsi^\ell_{(i)} - \bpsi^\ell) A_{2i-1,2i(-2i)} A_{2j-1,2i(-2i)}^{-1},
\end{equation}
which because of the identification \eqref{eq:aji-l} agrees with equation
\eqref{eq:L-ij}.
\begin{Rem}
The reverse identification from $K$ dimensional quadrilateral lattice $\psi$
and all quadrilateral lattices generated via the Laplace transformations
to the corresponding $2K-1$ Desargues lattice is based on the 
observation \cite{TQL}, that for the fixed direction $i$ of the 
quadrilateral lattice the
$2K$ points $\psi^\ell$, $\psi_{(i)}^\ell $, 
$\mathcal{L}_{ij}(\psi)^\ell$,
$\mathcal{L}_{ij}(\psi_{(j)}^\ell)$ are collinear. The corresponding lines (in
the present notation they are denoted $L_{(-2i)}$) form $i$-th tangent
congruence of the lattice $\psi^\ell$. 
\end{Rem}
\begin{Rem}
It is
known \cite{MQL} that $K$ dimensional quadrilateral lattice is uniquely
determined from a system of $K(K-1)/2$
quadrilateral surfaces intersecting along $K$
initial discrete curves which have one point in common. The successive
application of the Laplace
transformations generates then $2K-1$ dimensional Desargues lattice. Because a
quadrilateral surface is uniquely determined from two initial curves by two
functions of two discrete variables, therefore a solution of $2K-1$ dimensional
Hirota--Miwa
equation is determined given $K(K-1)$ functions of two (appropriate)
variables.
\end{Rem} 
\begin{Rem}
The Desargues lattices of even $N=2K$ dimension can be obtained as dimensional
reduction of $2K+1$ Desargues lattices (set $n_{2K+1}=0$). Equivalently, it is
generated by the Laplace transformations from a $K$ dimensional quadrilateral
lattice and focal lattices of a congruence conjugate to the lattice (see
\cite{TQL} for explanation of the terms used).
\end{Rem} 

\section{Conclusion and final remarks}
\label{sec:concl}
In the paper we studied an elementary geometric meaning of
the celebrated Hirota--Miwa system. The multidimensional compatibility of the
corresponding map relies on the Desargues theorem and its higher-dimensional
generalizations. Since the Desargues theorem is valid in projective spaces over
division rings, we are automatically led to the non-commutative
Hirota-Miwa system of equations. Notice, that the division ring context of the
Hirota--Miwa equation shouldn't be considered just
as a curiosity. It is
known \cite{Cliff,Panov,Jordan} that the standard quantum algebras 
\cite{Jimbo,Drinfeld,Woronowicz,RTF} admit division rings of quotients.  In view
of recent developments on quantization of the discrete Darboux
equations~\cite{BaSe,BaMaSe} this aspect of integrable discrete geometry deserves
deeper studies. 

Although the linear problem for the Desargues maps seems to be strong
degeneration of the linear problem for the quadrilateral lattice map,
surprisingly both theories are equivalent, as suggested by their equivalence
on the level of the algebro-geometric solutions, and those obtained by the
non-local $\bar\partial$-dressing method. We found also the
meaning of the $\tau$-function of the Hirota--Miwa equation for that
class of solution as a Fredholm determinant.

We would like to stress that the above-mentioned equivalence becomes
elementary and visible on the level of discrete systems. On the level of 
differential
equations the situation is much more subtle. It is however known~\cite{KoMA} 
that one component KP hierarchy can been reformulated, after the transition to
the so called Miwa
coordinates, as a system of infinite number of
(partial differential) Darboux equations. 

The theory of discrete integrable systems
is richer (see for example  \cite{Suris,DIS}) but also, in a sense,
simpler then the corresponding theory of integrable partial differential
equations. In the course of a limiting procedure, which gives differential 
systems from the discrete ones, various symmetries and relations between 
different discrete systems are lost or hidden. The present paper gives new 
example supporting this claim, and shows once again the superior role of the
(non-Abelian) Hirota--Miwa equation in the integrable
systems theory.

\section*{Acknowledgments}
I acknowledge discussions with Jaros{\l}aw Kosiorek and Andrzej
Matra\'{s} on the role of the
Desargues configuration in foundations of geometry, and an important warning by
Mark Pankov on a terminological confusion with the lattice theory.
I also would like to thank a Referee for his comments on the manuscript which
helped me to improve presentation. It is my pleasure to thank
the Isaac Newton Institute for Mathematical Sciences
for hospitality during the programme 
\emph{Discrete Integrable Systems}.
\bibliographystyle{amsplain}

\providecommand{\bysame}{\leavevmode\hbox to3em{\hrulefill}\thinspace}

\end{document}